\documentclass[floatfix,twocolumn,pre,showpacs]{revtex4}
\usepackage{graphicx}
\usepackage{psfrag}
\usepackage{epsfig}
\usepackage{subfigure}
\usepackage{amssymb,amsmath}
\usepackage{color}
 
\begin{document}
\title{Contact line dynamics in binary lattice Boltzmann simulations}
\author{C. M. Pooley, H. Kusumaatmaja and J. M. Yeomans}
\affiliation{The Rudolf Peierls Centre for Theoretical Physics, Oxford University, 1 Keble Road, Oxford OX1 3NP, U.K.}
\date{\today}
\begin{abstract}
We show that, when a single relaxation time lattice Boltzmann algorithm is used to solve the hydrodynamic equations of a binary fluid for which the two components have different viscosities, strong spurious velocities in the steady state lead to incorrect results for the equilibrium contact angle. We identify the origins of these spurious currents, and demonstrate how the results can be greatly improved by using a lattice Boltzmann method based on a multiple-relaxation-time algorithm. By considering capillary filling we describe the dependence of the advancing contact angle on the interface velocity.
\end{abstract}
\maketitle


\section{Introduction}
 
There are many different physical systems for which contact line dynamics plays an important role. These range from large scale industrial processes, such as oil recovery and coating, to micron scale phenomena, such as flow in microfluidic channels or the movement of drops across micropatterned surfaces. Despite its importance, contact line dynamics is still a challenging problem that has not been fully solved. This is partly because it is an inherently multiscale problem. The hydrodynamic flow of the fluids far away from the three phase contact line has to be matched to the microscopic details of the contact line motion \cite{cox}. In particular, a mechanism is needed by which the usual hydrodynamic, no-slip boundary condition can be violated in the vicinity of the contact line \cite{huh, qian1}. 

One solution to this problem has been to consider a class of hydrodynamic models, called diffuse interface models \cite{jacqmin, seppecher, briant}, where the interface at the contact line has a finite width. As a diffuse interface is pushed across a surface it is pulled out of equilibrium. This leads to diffusive transport of fluid across the interface and hence an effective slip at the contact line. Here we shall consider a diffuse interface model for a binary fluid, where the equilibrium properties of the fluid are described by a Landau free energy functional. The equilibrium contact angle is controlled by a surface term in the free energy.

The hydrodynamic equations of motion of diffuse interface models can be solved in many different ways, but in this paper, we shall focus on one particular mesoscopic modelling technique, the lattice Boltzmann method \cite{swift, Benzi, succibook}. Lattice Boltzmann simulations have been succesfully used to study several contact line problems. Examples include the spreading of drops on chemically patterned \cite{leopoldes} and superhydrophobic surfaces \cite{dupuis}, modelling of contact angle hysteresis \cite{halimhysteresis}, and capillary filling in microfluidic channels \cite{influs, succi1, Timonen}. 

The majority of previous work on binary fluids has concentrated on the case where the two components have equal viscosity. In this case a simple and widely used lattice Boltzmann approach, the BGK algorithm, works well, agreeing with analytical results for the equilibrium contact angle. However, if the two binary components have different viscosities, this is no longer the case: we shall show that the equilibrium contact angle is predicted incorrectly by the algorithm because there are severe spurious velocities in the steady state.

There are several important aspects of the behaviour of multicomponent fluids where it is essential or highly desirable to be able to model a two component fluid, where the components have different viscosities. These include drops moving on surfaces, where the viscosity of the surrounding fluid must be substantially smaller than that of the drop to access a rolling regime, instabilities such as viscous fingering which are driven by a viscosity difference between the two fluid components, and capillary filling, where the simple theories assume that the displaced fluid has zero viscosity.  

In this paper we identify two primary reasons for the spurious currents in BGK lattice Boltzmann simulations of contact line hydrodynamics. We show how the spurious effects can be greatly suppressed by using an algorithm based on a multiple-relaxation-time lattice Boltzmann approach. We demonstrate that the new algorithm gives excellent agreement with theory, both for the equilibrium contact angle, and for the advancing contact angle, measured in capillary filling simulations.

The paper is organised in the following way: We begin, in Sec.~\ref{secbin}, by introducing the free energy model for a binary fluid system. We summarise two different lattice Boltzmann implementations that can be used to solve the binary fluid's hydrodynamic equations of motion, the BGK model and a multiple-relaxation-time method, in Secs.~\ref{latimp}A and B. In Sec.~\ref{eqcontangle}, we measure the equilibrium contact angle for a drop on a surface and find that, for the BGK approach, it deviates from the value predicted theoretically if there is a viscosity ratio between the two phases. This deviation is caused by anomalous spurious currents near to contact points. In Sec.~\ref{spurious}, we discuss the origin of the spurious currents: firstly long ranged effects and secondly non-zero velocities induced by the bounce-back boundary conditions. We propose, in Sec.~\ref{algorithm}, an algorithm based on a multiple-relaxation-time lattice Boltzmann implementation \cite{HG02,PA06} which significantly suppresses the spurious currents at the contact point. In Sec.~\ref{washburn} we use simulations of capillary filling in smooth channels to measure the dependence of the advancing contact angle on capillary number. Finally, in Sec.~\ref{summary}, we summarise the results and conclude.

\section{Modelling the Binary Fluid}
\label{secbin}

The equilibrium properties of a binary fluid can be described by a Landau free energy functional \cite{briant}
\begin{eqnarray}
F = \int \{\psi + \frac{\kappa}{2} \left| \nabla \phi \right|^2\} dV + \int  \{ \xi \phi \} dS,
\label{freeen}
\end{eqnarray}
where the bulk free energy density $\psi$ is taken to have the form
\begin{eqnarray}
\psi = \tfrac{c^2}{3} \rho \ln \rho + a \left( -\tfrac{1}{2} \phi^2 + \tfrac{1}{4} \phi^4 \right).
\label{bulkfe}
\end{eqnarray}
The first integral in Eq.~(\ref{freeen}) is taken over the volume of the system. $\rho$ is the fluid
density and $\phi$ is the concentration. $c$ is a lattice velocity parameter, described below and $a$ is a constant.
This choice of $\psi$ gives binary phase separation into two phases with $\phi = \pm 1$. The $\kappa$  
term in Eq.~(\ref{freeen}) represents an energy contribution from gradients in $\phi$ and is related to  
the surface tension between the two phases through $\gamma = \sqrt{8\kappa a/9}$ \cite{briant}.

The second integral in Eq.~(\ref{freeen}) is over the system's surface and is used to model the fluid-solid 
surface tensions \cite{Cahn}. The parameter $\xi$ determines the contact angle. 

Taking the functional derivative of Eq. (\ref{freeen}) with respect to $\phi$ gives the chemical potential 
\begin{eqnarray}
\mu = \tfrac{\delta F}{\delta \phi} = a\left(- \phi + \phi^3 \right)- \kappa \nabla^2 \phi,
\label{chempot}
\end{eqnarray} 
which is constant in equilibrium. Minimisation of the free energy also shows that the gradient in $\phi$ at the 
boundary is $\left .\partial_\perp \phi \right|_b =  {\xi}/{\kappa}$ \cite{briant}, where the partial derivative $\partial_\perp$ 
is taken in a direction normal to  the surface.

The dynamics describing how the fluid approaches equilibrium are determined by the pressure  
tensor
\begin{eqnarray}
P_{\alpha \beta} =  \left(p_0 - \kappa \phi \nabla^2 \phi - \tfrac \kappa 2 |\nabla \phi|^2  
\right) \delta_{\alpha \beta} + \kappa \partial_{\alpha} \phi  \, \partial_{\beta} \phi,
\label{pressten}
\end{eqnarray}
where the bulk pressure is
\begin{eqnarray}
p_0 =  \tfrac{c^2}{3}\rho  +  a \left( -\tfrac{1}{2} \phi^2 + \tfrac{3}{4} \phi^4 \right).
\label{p0}
\end{eqnarray}

The hydrodynamic equations for the binary fluid are \cite{briant}   
\begin{eqnarray}
 \partial_{t} \rho + \partial_{\alpha} (\rho v_{\alpha}) &=& 0, \label{nsfinal21} \\
 \partial_{t}(\rho v_\beta) +  \partial_\alpha ( \rho v_\alpha v_\beta) &=& -  \partial_\alpha 
P_{\alpha \beta} \label{nsfinal22} \\
&&+ \partial_\alpha  \left\{  \nu \rho \left( \partial_\beta  v_\alpha + 
\partial_\alpha  v_\beta \right)  \right\},\nonumber\\
 \partial_t{\phi} + \partial_\alpha \left( \phi v_\alpha \right) &=& M \nabla^2 \mu
\label{nsfinal23}
\end{eqnarray}
where ${\bf v}$ is the fluid velocity, $\nu$ is the kinematic viscosity and $M$ is a mobility. 
The equilibrium properties of the fluid appear in the equations of motion through the pressure
tensor and the chemical potential.

\section{Lattice Boltzmann Implementations}
\label{latimp}

The equations of motion \eqref{nsfinal21}-\eqref{nsfinal23} can be solved using a lattice Boltzmann algorithm. We shall consider two different lattice Boltzmann approaches in this paper and, for convenience, we summarise both here.

\subsection{Single-relaxation-time lattice Boltzmann}

To implement a lattice Boltzmann algorithm for a binary fluid in two dimensions the system is divided up into a square grid of points, with two particle distribution functions $f_i({\bf r}, t)$ and $g_i({\bf r}, t)$ on each point. The label $i$ denotes a particular lattice velocity vector ${\bf e}_i$, defined by ${\bf e}_0 = (0,0)$, ${\bf e}_{1,2} = (\pm c,0)$, ${\bf e}_{3,4} = (0,\pm c)$, ${\bf e}_{5,6} = (\pm c,\pm c)$, and ${\bf e}_{7,8} = (\mp c,\pm c)$. The lattice velocity parameter $c$ is defined by $c = {\Delta x}/{\Delta t}$, where $\Delta x$ is the spacing between nearest neighbouring points and $\Delta t$ is the time step. The physical variables are obtained from the particle distribution functions using
\begin{eqnarray}
\rho  = \sum_i f_i, \quad \phi = \sum_i g_i, \quad \rho {\bf v} =  \sum_i f_i {\bf e}_i.
\label{moments}
\end{eqnarray}

The time evolution equation for the particle distribution functions, using the standard BGK approximation \cite{succibook}, can be broken down into two steps. The first is a collision step
\begin{eqnarray}
f_i^\prime({\bf r}, t) &=& f_i({\bf r}, t) - \tfrac{1}{\tau_\rho} \left[  
f_i - f_i^{eq} \right], \label{latbolt} \\
g_i^\prime({\bf r}, t) &=& g_i({\bf r}, t) - \tfrac{1}{\tau_\phi} \left[  
g_i - g_i^{eq} \right]. \label{gevo}
 \label{latbolt2}
\end{eqnarray}
This is followed by a streaming step, which moves particles along their corresponding lattice velocity vector direction
\begin{eqnarray}
f_i({\bf r} + {\bf e}_i \Delta t , t+\Delta t) &=& f_i^\prime({\bf r}, t), \label{stream1} \\
g_i({\bf r} + {\bf e}_i \Delta t , t+\Delta t) &=& g_i^\prime({\bf r}, t). \label{stream2}
\end{eqnarray}
$f^{eq}_i$ and  $g^{eq}_i$ are equilibrium distribution functions, defined as a power series in the velocity.
A summary of our choice of equilibria, which is motivated to help reduce spurious velocities around  
interfaces \cite{kalli}, is given in Appendix \ref{app1}. Note, the inter-coupling between $f_i$ and $g_i$ comes through $f^{eq}_i$ and  $g^{eq}_i$. In particular, the large variation in $\phi$ at an interface influences $f_i$ by $\partial_\alpha \phi$ and $\nabla^2 \phi$ terms in $f_i^{eq}$.

A Chapman Enskog expansion can be performed to show that the lattice Boltzmann collision (\ref{latbolt}, \ref{latbolt2}) and streaming (\ref{stream1}, \ref{stream2}) operations lead to the hydrodynamic equations \eqref{nsfinal21}-\eqref{nsfinal23} in the limit of long length and time scales \cite{swift}. The relaxation parameters $\tau_{\rho}$ and $\tau_{\phi}$ are related to the kinematic viscosity and mobility through
\begin{eqnarray}
\nu = \Delta t \tfrac{c^2}{3} \left( \tau_\rho - \tfrac{1}{2} \right)\label{kine} , \\
M =  \Delta t \Gamma \left(\tau_\phi - \tfrac{1}{2}\right)\label{Mobility} ,
\end{eqnarray}
where $\Gamma$ is a tunable parameter that appears in the equilibrium distribution.

\subsection{Multiple-relaxation-time lattice Boltzmann}
\label{MRT}

We next summarise the basic methodology behind the multiple-relaxation-time (MRT) lattice Boltzmann approach. More details are given in \cite{HG02,PA06}. The idea behind multiple-relaxation-time lattice Boltzmann is that different relaxation parameters are used for different linear combinations of the distribution functions. In particular, the relaxation parameters responsible for generating the viscous terms in the Navier-Stokes equation \eqref{nsfinal22} are set to $\tau_{\rho}$, those connected to conserved quantities to $\infty$, and all others to $1$. 

In multiple-relaxation-time lattice Boltzmann the collision term $\tfrac{1}{\tau_\rho} \left[ f_i - f_i^{eq} \right]$ on the right hand side of the lattice Boltzmann equation for $f_i$ (\ref{latbolt}) is replaced by
\begin{eqnarray}
{\bf M}^{-1} {\bf S} {\bf M} \left[ {\bf f} - {\bf f}^{eq} \right], 
\label{mrtlb}
\end{eqnarray}
where the particle distributions $f_i$ and $f_i^{eq}$ are written as column vectors and ${\bf M}$ is a matrix. One possible choice for ${\bf  M}$ is \cite{DS06}
\begin{eqnarray}
{\bf M} = \left(
\begin{array}{ccccccccc}
1 & 1 & 1 & 1 & 1 & 1 & 1 & 1 & 1\\
-4 & -1 & -1 & -1 & -1 & 2 & 2 & 2 & 2\\ 
4 & -2 & -2 & -2 & -2 & 1 & 1 & 1 & 1\\
0 & 1 & -1 & 0 & 0 & 1 & -1 & -1 & 1 \\
0 & -2 & 2 & 0 & 0 & 1 & -1 & -1 & 1 \\
0 & 0 & 0 & 1 & -1 & 1 & -1 & 1 & -1 \\
0 & 0 & 0 & -2 & 2 & 1 & -1 & 1 & -1 \\
0 & 1 & 1 & -1 & -1 & 0 & 0 & 0 & 0 \\
0 & 0 & 0 & 0 & 0 & 1 & 1 & -1 & -1 \\
\end{array}
\right).
\label{Matrix}
\end{eqnarray}
This matrix performs a change of basis. The new basis is designed to contain more physically  
relevant variables. For instance, when the first row is dotted with ${\bf f}$ the density $\rho = \sum_i f_i$ is  
obtained. Similarly, the fourth and sixth lines calculate the momentum densities $\rho u_x$ and $\rho u_y$,  
respectively. Each of the rows in ${\bf M}$ is mutually orthogonal so the inverse follows easily as 
\begin{eqnarray}
{\bf M}_{ij}^{-1} = \frac{1}{\sum_k {\bf M}^2_{jk}}{\bf M}_{ji}.
\end{eqnarray}

The matrix ${\bf S}$ in Eq. \eqref{mrtlb} is diagonal and has the elements
\begin{eqnarray}
{\bf S} = \text{diag} \left(0,1,1,0,1,0,1, \omega,\omega \right)
\label{diag}
\end{eqnarray}
where $\omega = 1/ \tau_\rho$ now determines the fluid viscosity in Eq. \eqref{kine}. Note that  
some of the values are zero. This choice is arbitrary as these modes correspond to the conserved  
moments; $\sum_i M_{ji} \left[ f_i -  f^{eq}_i \right] = 0$ for $j = 0,3,5$. The choice of unity for the
other terms will be justified later.

It is not necessary to adopt a multi-relaxation approach for $g_i$ as there is an independent parameter $\Gamma$, which can be varied to change the mobility of particles in Eq. (\ref{Mobility}). Therefore, even when using a MRT approach, we set distribution $g_i$ to evolve according to Eq. (\ref{gevo}) with $\tau_\phi = 1$.

\section{Measuring the Equilibrium Contact Angle}
\label{eqcontangle}   

In this section we check the extent to which the single relaxation time lattice Boltzmann method gives the correct equilibrium contact angle at the contact point. We find that, for a fluid with a constant viscosity modelled using $\tau_\rho=\tau_{\alpha}=\tau_{\beta}=1$, where $\alpha$ and $\beta$ label the two coexisting bulk phases, good results are obtained. However, when we consider a difference in viscosity between the two components, $\tau_{\alpha}=3$ and $\tau_{\beta}=0.7$, the approach does not work well. In the next section we shall explain why not and describe a method to overcome the problem.

\begin{figure}
\begin{center}
\includegraphics[width = 3.in]{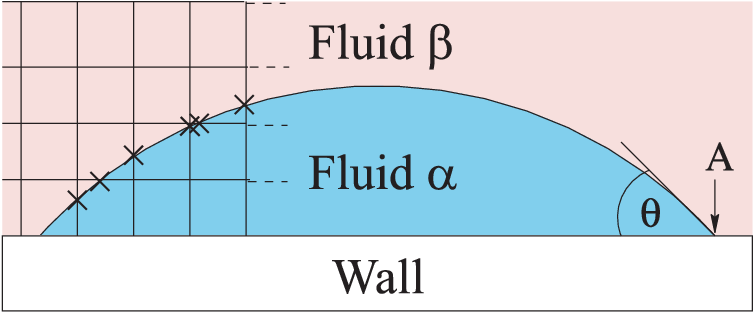}
\end{center}
\caption{(Color online) The simulation geometry. The contact angle is defined by $\theta^{eq}$. The crosses denote the points which are 
fit to the arc of a circle to obtain a numerical estimate of the equilibrium contact angle $\theta^{eq}$. The grid is not
to scale.}
\label{fig7}
\end{figure}
Fig.~\ref{fig7} shows a drop resting on a solid surface. In general, the fluid-solid ($\gamma_{\alpha s}$ and $\gamma_{\beta s}$) and fluid-fluid ($\gamma$) surface tensions are different. At the contact point $A$ the balance of forces is described by Young's law
\begin{eqnarray}
\cos \theta^{eq} = \frac{\gamma_{\beta s} - \gamma_{\alpha s}}{\gamma},
\label{Young}
\end{eqnarray}
where $\theta^{eq}$ defines the equilibrium contact angle.

We performed simulations to verify this relation. The lattice Boltzmann system size was set to $300 \times 100$ lattice units and the parameters used were  $a = 0.04$, $\tau_\phi = 1$, $\Gamma  
= 0.5$, $\Delta t = \Delta x = 1$, and $\kappa = 0.04$, giving an interface width of $W = 2\sqrt{2\kappa/a} = 2.8$ lattice sites and an interfacial tension of $\gamma = 0.038$ in lattice units.

The relaxation parameter $\tau_\rho$ was determined by
\begin{eqnarray}
 \tau_\rho = \tau_{\beta} + \tfrac{ \phi + 1 }{2} \left( \tau_{\alpha} - \tau_{\beta} \right)
\label{tau}
\end{eqnarray}
such that it changed smoothly through the interface and had the bulk values $\tau_{\alpha}$ and $\tau_{\beta}$ in the two bulk phases. The values $\tau_{\alpha} = 3$ and $\tau_{\beta} = 0.7$ were chosen to give a viscosity ratio of $R_\nu = ({\tau_{\alpha}-0.5})/({\tau_{\beta}-0.5}) = 12.5$.

Non-slip boundary conditions at the walls were achieved using a standard mid-link, bounce-back method \cite{LV00}. The contact angle was varied by changing the gradient of the order parameter at the solid boundary, $\left. \partial_\perp \phi \right|_b$. Initially, a semi-circle of fluid of radius $R = 35$  was placed on the surface with a contact angle of $90^\circ$. The boundary conditions were set to $\left. \partial_\perp \phi \right|_b = 0$ and the system was evolved for $3\times10^4$ time-steps, such that it reached equilibrium. $\left. \partial_\perp \phi \right|_b$ was then quasi-statically increased over the course of $10^6$ time steps and the variation in the contact angle was measured. The process was repeated, but this time decreasing $\left. \partial_\perp \phi \right|_b$ until the surface was completely wet.  

Numerical measurements of contact angle were performed by matching the interface to the arc of a circle \cite{Rothman}.
Specifically, each link between neighbouring lattice sites was examined to see if at one end it was fluid $\alpha$ ($\phi > 0$) and at the other end fluid $\beta$ ($\phi < 0$). If this was the case then a linear interpolation was used to predict the point on the link where $\phi = 0$. These crossing points are illustrated by the crosses in Fig.~\ref{fig7} (note that this grid is not to scale). The equilibrium contact angle was estimated by performing a least squares fit of the crossing points to a circular section and then calculating the contact angle the section made with the surface. 

\begin{figure}
\begin{center}
\includegraphics[width = 3.3in]{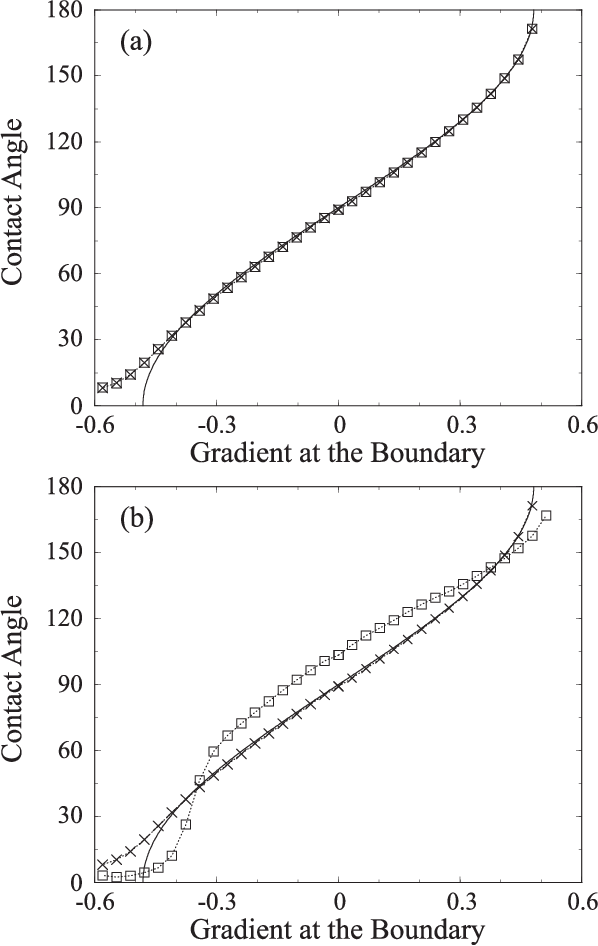}
\end{center}
\caption{The equilibrium contact angle as a function of the gradient in $\phi$ at the boundary (measured in lattice units), $\left. \partial_\perp \phi \right|_b$, for (a) $\tau_{\alpha} = \tau_{\beta} = 1.0$ and (b) $\tau_{\alpha} = 3$, $\tau_{\beta} = 0.7$. Squares and crosses are simulation results obtained using single and multiple-relaxation-time lattice Boltzmann algorithms respectively. The solid curve is the theoretical expression \eqref{theory}.}
\label{fig2}
\end{figure}
%

Results for the equilibrium contact angle for different values of $\tau_{\alpha}$ and $\tau_{\beta}$ are shown in Fig.~\ref{fig2} and compared to the exact result \cite{briant},
\begin{eqnarray}
\sqrt{\frac{2 \kappa}{ a}} \left. \partial_\perp \phi \right|_b &=& 2 \,\text{sgn} \left(\theta^{eq} \!- \! \frac{\pi}{2} \right)  \nonumber \\
&& \times \left[ \cos\left(\frac{\alpha}{3} \right) \left\{ 1 \!- \!  
\cos\left(\frac{\alpha}{3}\right)\right\} \right]^{\tfrac{1}{2}},
\label{theory}
\end{eqnarray}
where $\alpha = \arccos \left(\sin^2 \theta^{eq} \right)$. The agreement is good for $\tau_{\alpha} = \tau_{\beta} = 1$ (Fig.~\ref{fig2}(a)), but there are large errors for $\tau_{\alpha} \neq \tau_{\beta}$ (Fig.~\ref{fig2}(b)). In the next section we discuss the reasons behind this discrepancy.  

\section{The Origins of Spurious Currents Near the Contact Point}
\label{spurious}

\begin{figure}  
\begin{center}
\includegraphics[width = 3.in]{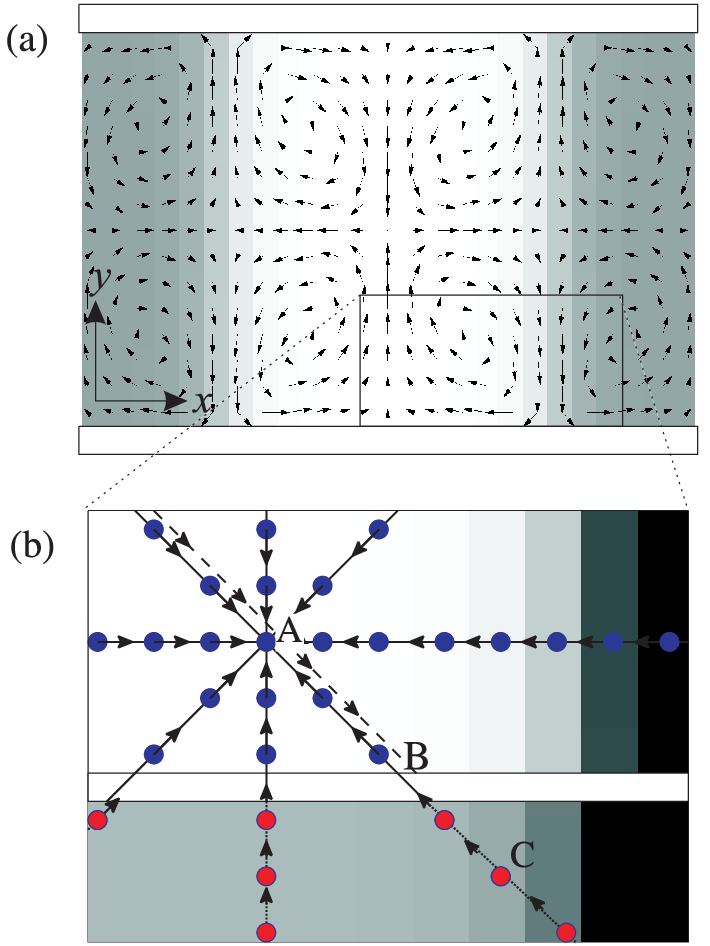}
\end{center}
\caption{(a) The spurious flow field in steady state in a system consisting of a stripe of one fluid (light region) surrounded by another fluid (dark regions) held between two walls. Periodic boundary conditions are applied in the $x$-direction. The parameters used were $\tau_{\alpha} = \tau_{\beta} = 10$. (b) (Color online) An enlargement of the region near the interface indicating how the particle distribution function at A depends on contributions from lattice nodes in the lattice velocity vector directions.}
\label{fig1}
\end{figure}

The reason that the single relaxation time lattice Boltzmann approach gives an incorrect equilibrium contact angle for $\tau_{\rho} \neq 1$ is that near to the contact point there are strong spurious velocities which continuously push the system out of equilibrium and result in the deformation of the drop. Note that even when the surface is neutrally wetting, $\left. \partial_\perp \phi \right|_b = 0$, the wetting angle measured numerically is $\sim 10^\circ$ larger than the expected value of $\theta^{eq}=90^\circ$. We focus on this case and replace the drop geometry with a simpler system consisting of a stripe of component $\alpha$ between two neutrally wetting walls, as depicted in Fig.~\ref{fig1}(a).
This simulation was performed for a system of size $28\times15$ lattice units using a large  
relaxation parameter $\tau_\rho = \tau_{\alpha} = \tau_{\beta} = 10$. Note that we do observe the correct  
$90^\circ$ contact angle in this case, but only because the viscosities of the two phases 
are the same. If we set $\tau_{\alpha} > \tau_{\beta}$ then the stripe becomes barrel shaped,  
consistent with the measurements reported in Fig.~\ref{fig2}(a). 

The black arrows in Fig. \ref{fig1}(a) show the steady-state, spurious velocity field in the system. The magnitude of the velocities is typically of order $10^{-3}c$. We have identified two contributions to the spurious velocities; one from long range effects, a second from the bounce back boundary conditions. We discuss each in turn.

\subsection{Spurious velocities from long range effects}
\label{longrange}

For a unbounded, system at steady-state, it is possible, by iterating the lattice Boltzmann evolution equation 
(\ref{latbolt}), to write the particle distribution function at any given lattice point in terms 
of equilibrium distributions along lines defined by the lattice velocity vector directions:
\begin{eqnarray}
f_i({\bf r}) = \sum_{j=1}^\infty \frac{1}{\tau_\rho} \left( 1-\frac{1}{\tau_\rho} \right)^{j-1}  
f^{eq}_i ({\bf r} - j \Delta t {\bf e}_i).  
\label{ss}
\end{eqnarray}
Note that when $\tau_\rho = 1$ this reduces to $ f_i({\bf r}) =  f^{eq}_i ({\bf  
r} -  \Delta t {\bf e}_i)$. This special case could be described as `local', in the sense that the particle  
distribution function only depends on the equilibria of its neighbours. 

When $\tau_\rho > 1$, contributions to the sum in Eq.~(\ref{ss})  decay with a characteristic length  
\begin{eqnarray}
\lambda_{\tau_\rho>1} = \left[ \ln \left( \tfrac{\tau_\rho}{\tau_\rho-1} \right) \right]^{-1}\Delta x.
\label{decay}
\end{eqnarray}
This diverges as $\tau_\rho \rightarrow \infty$. When $\tau_\rho < 1$, each term in the sum  
(\ref{ss}) alternates in sign and the envelope decays exponentially with length scale
\begin{eqnarray}
\lambda_{\tau_\rho<1}  = \left[ \ln \left( \tfrac{\tau_\rho}{ 1-\tau_\rho} \right) \right]^{-1} \Delta x
\label{decay2}
\end{eqnarray}
which diverges as $\tau_\rho \rightarrow \tfrac{1}{2}$ (This limit makes sense because it corresponds  
to the viscosity in Eq.~(\ref{kine}) becoming zero.) 
Therefore, for $\tau_\rho$ both high and low, the distribution at any given point is dependent on other nodes a long distance away.

Even without the presence of solid boundaries, it has been noted that these long range interactions 
can give rise to large spurious currents around curved interfaces \cite{kalli}. In particular, the size of 
the spurious velocities scales as $\tau_\rho^3$ for large $\tau_\rho$, so choosing $\tau_\rho$ large is not advisable. The problem
becomes even more acute in the presence of solid boundaries.    

Figure \ref{fig1}(b) schematically shows the situation for a particular lattice node $A$ close to the contact point. Because we focus on the case of neutrally wetting walls (contact angle $90^o$), this allows us to compare two systems we expect to have the same steady state configuration, {\it i.e.} a rectangular stripe of fluid at rest. In the first, simpler system we imagine that the walls have been replaced by periodic boundary conditions in the $y$-direction. In this case, contributions to Eq.~(\ref{ss}) radiate out indefinitely in all $8$ directions. This is illustrated in Fig. \ref{fig1}(b) by the filled circles, each of which denote a term in the sum on the right hand side of Eq.~(\ref{ss}). (Note, the arrows indicate the direction of motion for the distribution function from which each term originates.) Because the boundaries have been removed, the symmetry of the system implies that any point above or below $A$ at the same value of $x$ must behave exactly the same as $A$. Since there can be no net flux of fluid across any given plane (as this would lead to non-conservation of mass), then in steady state, the system must be at rest. It can be numerically confirmed that no steady state spurious velocities are generated in this case.

In the second system, the periodic boundary conditions are replaced with bounce back boundary conditions, and we now discuss what implications this has. In particular, Eq.~(\ref{ss}) needs to be modified to take into account the fact that the contributions to the sum do not extend indefinitely in all $8$ directions because of the presence of the boundaries. This is illustrated at point $B$ in Fig. \ref{fig1}(b), where the remaining terms in the sum, which would have originated in the direction of $C$, come, in fact, from a reflected branch in the direction of $A$. Because of the interface in the system, these two paths give significatly different contributions to the sum. Therefore, the reflection of branches at the boundary breaks the symmetry of the system in the $y$ direction. (This can clearly be seen if we compare point A to a point on the boundary, where the incoming branches are reflected immediately.) This broken symmetry leads to the generation of the spurious, steady state velocities.

The range around the contact point over which this spurious force is active is determined by the  
decay lengths $\lambda$ in Eqs.~(\ref{decay}) and (\ref{decay2}). For high or low   
viscosities $\lambda$ is large and this is a long range effect. Correcting for it  
at the boundary involves extrapolating into the surface to predict, for example, the density  
variation down the dotted branches near to $C$. 
This is relatively simple for the $90^\circ$ case shown, as the unknown nodes can be obtained by  
reflecting the system across the boundary. However, in the case of arbitrary contact angle the  
situation is much more complicated and a general solution is far from clear.  

Thus implementing solid boundaries in the multi-component BGK lattice Boltzmann is  
only advisable in the case $\tau_\rho = 1$. (For the same reason it is best to choose $\tau_\phi = 1$ \cite{tau}.)
However, to simulate phases with different viscosities it is not possible to use $\tau_\rho = 1$ in both phases. 
In Sec.~\ref{algorithm} we discuss how using a multiple-relaxation-time method can be used to overcome this restriction.   

\subsection{Spurious velocities from bounce-back boundary conditions}

\begin{figure}
\begin{center}
\includegraphics[width = 2.6in]{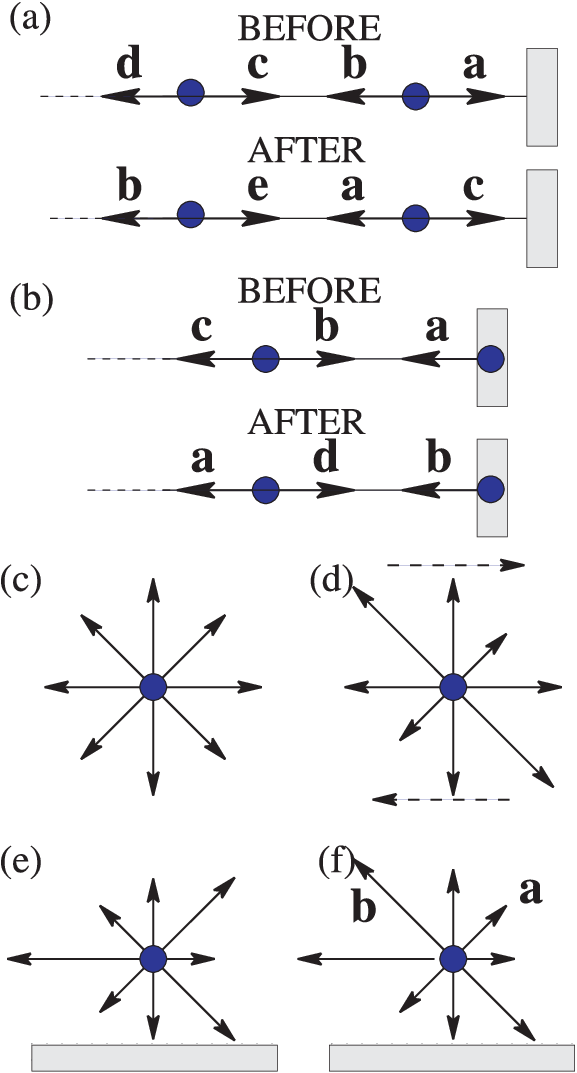}
\end{center}
\caption{(Color online) (a) The mid-link bounce-back method, before and after the streaming step, along one  
lattice direction. (b) The lattice node bounce-back method. (c-f) Schematic diagrams showing the  
particle distribution functions for (c) a single-component fluid, (d) a single-component fluid under shear, 
(e) particle distribution function in the interface of a binary fluid before and (f) after bounce back.}
\label{fig9}
\end{figure}

A second source of spurious velocities at the contact point is the bounce-back boundary conditions.
To understand what goes wrong with the boundary conditions for the binary lattice Boltzmann  
model it is first important to understand why they work well when simulating a single-component fluid. 
Fig.~\ref{fig9}(a) illustrates the mid-link, bounce-back process for one dimension. During the streaming the  
population {\bf a} moves down the link, is reflected, and then travels back to the node it came from but  
moving in the opposite direction. (Fig.~\ref{fig9}(b) shows an alternative bounce-back scheme in which the wall lies on the boundary nodes. 
The arguments presented here are equally valid for this case.)

Fig.~\ref{fig9}(c) shows the distribution for a typical node in a system describing a single component fluid at equilibrium. This is a  
representation of the Maxwell Boltzmann distribution using a discrete number of velocity vectors.
The bounce-back method acts at the boundaries of the system to switch some of the populations to travel in the opposite direction.
 Because this distribution is invariant under a parity transformation (that is replacing velocity vectors ${\bf v}$  
by ${-\bf v}$) then the correct distribution is preserved by the bounce-back boundary. When the  
system is in a state of shear the particle distribution function, in the rest frame of the fluid, is as depicted in Fig.  
\ref{fig9}(d). While this is clearly not isotropic, it still preserves invariance under the parity  
transformation and hence the bounce-back approach is still valid. (In fact, if non-slip boundary conditions are
enforced by setting the particle distribution at the boundary to its equilibrium value at rest,  
an inaccurate shear velocity profile is obtained. This is because the shear-induced distortion of the distribution function 
is not preserved at the boundaries.)

We now return to the binary fluid case. Fig.~\ref{fig9}(e) shows the distribution function for a typical node, lying at rest in a fluid-fluid interface. (In this particular example the interface lies perpendicular to the surface, hence the distribution has an up-down symmetry.) Note that because of the $\kappa$ terms in the pressure tensor (\ref{pressten}) the parity invariance is broken. We consider the case when the position of the node is $\Delta x/2$ above a boundary and mid-link bounce-back boundary conditions are being employed. Fig. \ref{fig9}(f) illustrates the situation after the bounce back step, and it clearly shows that the new distribution is not the same as in (e) (in particular, vector ${\bf a}$ is shortened and ${\bf b}$ lengthened). Therefore, bounce-back collisions result in the system continually being pushed out of equilibrium, leading to the generation of spurious velocities on or near to boundaries close to the interface between the two phases.    

For $\tau_\rho = 1$, the distribution function is automatically set to its equilibrium value at each time step. While the typical distribution function in the interface is not invariant under a parity transformation, the equilibrium distribution is. Hence the spurious velocities caused by bounce back boundary conditions are suppressed in this case.
 
\section{Suppressing the spurious currents}
\label{algorithm}

We now describe how an algorithm based on the multiple-relaxation-time lattice Boltzmann 
approach can be used to significantly reduce the spurious currents at the contact point.
The approach comprises the following four steps: 

{\bf Step 1:} Calculate the density, concentration and velocity using the moments defined by Eq.~(\ref{moments}). 

{\bf Step 2:} Set the velocity of boundary nodes to zero when calculating the equilibrium distribution function.

{\bf Step 3:} Use the  multiple-relaxation-time lattice Boltzmann method described in Sec.~\ref{MRT} to perform the collision step.
	
{\bf Step 4:} Perform the streaming step with bounce-back at the boundaries.

To justify this choice we need to consider the hydrodynamic and the non-hydrodynamic modes separately.
From a hydrodynamic point of view, a non-slip boundary fixes the velocity at that boundary to zero. If the fluid is incompressible, then near to the boundary the fluid flow profile can always be approximated by a shear profile, with streamlines parallel to the surface. In section \ref{spurious}B we argued that for a shear profile, the hydrodynamic modes, which are represented by the difference between Figs. \ref{fig9}(c) and (d), are invariant under a parity transformation, and so are preserved by the bounce-back method. Another interpretation of this is that, for the hydrodynamic modes, the contributions at A in Fig. \ref{fig1} from the reflected branches are the same as those when periodic boundary conditions are considered (provided $\lambda_{\tau_\rho}$ is smaller than the length scale over which the shear profile approximation breaks down as we move away from the boundary). In summary, the hydrodynamic modes behave correctly for both the single and multiple-relaxation-time lattice Boltzmann models. The difference lies in the treatment of the non-hydrodynamic modes. 

In the multiple-relaxation-time scheme, because we have chosen $\tau_\rho = 1$ for the non-hydrodynamic modes (this choice corresponds to the 1's in Eq. (\ref{diag})), the spurious velocities generated from long ranged effects in the bulk are immediately removed. This justifies the use of the multiple-relaxation-time algorithm in Step 3. The only potential problem that remains is on the boundary nodes themselves. As discussed at the end of section \ref{spurious}B, the distribution function in the interface between fluid phases is not invariant under a parity transformation. This generates spurious velocities at any boundary node in an interface immediately after the streaming step 4. This problem is remedied by the introduction of step 2. Note that this step is consistent with non-slip boundary conditions and does not affect the hydrodynamic correlations (i.e. the non symmetric distribution in Fig. \ref{fig9}(d)).     
 


For a system with variable viscosity it would seem necessary to recalculate the collision matrix ${\bf C} = {\bf M}^{-1} {\bf S} {\bf M}$ in \eqref{Matrix} at each lattice node and at each time-step. This would be extremely slow computationally and not very practical. The approach we take is to create a lookup table with $\sim 10^4$ different values of viscosity and simply pick the closest match. 
	  
We find that implementing multiple-relaxation-time lattice Boltzmann with appropriate boundary conditions leads to a significant  
improvement in the accuracy of the equilibrium contact angle. The results of simulations for $\tau_{\alpha}=3$, $\tau_{\beta}=0.7$, are denoted by the crosses in Fig. \ref{fig2}(a), and show very good agreement with the theoretical, dashed curve. Deviation is only noticeable at small contact angles. This happens for two reasons: Firstly, the dynamics of drop wetting become very slow as the equilibrium contact angle becomes small, and so the assumption of quasi-static equilibrium, as the gradient in $\phi$ is slowly reduced, breaks down. (Tanner's law \cite{dupuis3} states that for a completely wetting surface in two dimensions, the size of a drop spreading on the surface scales as $r \sim t^{1/7}$ in time.) Secondly, the finite width of the interface $W$, which is neglected when assuming that the drop should be made up of a circular section, becomes comparable to the height of the drop. 
 
\section{A dynamical test: capillary filling}
\label{washburn}

We have shown that the equilibrium contact angle is accurately recovered
in lattice Boltzmann simulations for a binary system with different viscosities only when
a multiple-relaxation-time approach  is employed. While this highlights a 
problem with the single relaxation time binary lattice Boltzmann
model, contact angle measurement is a static problem. In this section we concentrate on the 
dynamics of a fluid penetrating a smooth microchannel to measure how the advancing contact 
angle changes with the velocity of the fluid interface.  

Fluid is pulled into a hydrophilic capillary by the Laplace pressure across the interface. Balancing this against the viscous drag
of the fluid column gives Washburn's law \cite{Washburn} describing the variation of the length $l$ of fluid in the capillary with time $t$
\begin{eqnarray}
l^2=\left( \frac{\gamma h \cos{\theta^a}}{3\rho \nu} \right) \, \left({t+t_0}\right),
\label{washburnlaw} 
\end{eqnarray}
where $h$ is the width of the capillary, $\theta^a$ is the advancing contact angle  and
$t_0$ is an integration constant. Note that it is appropriate to use, not the  
static, but the advancing contact angle  $\theta^a$, as this controls the curvature of the  
interface and hence the Laplace pressure. Moreover, Eq.~(\ref{washburnlaw}) assumes that there is no resistance 
to motion from any fluid already in the capillary. 

\begin{figure*}[ht]
\begin{center}
\includegraphics[scale = 0.85]{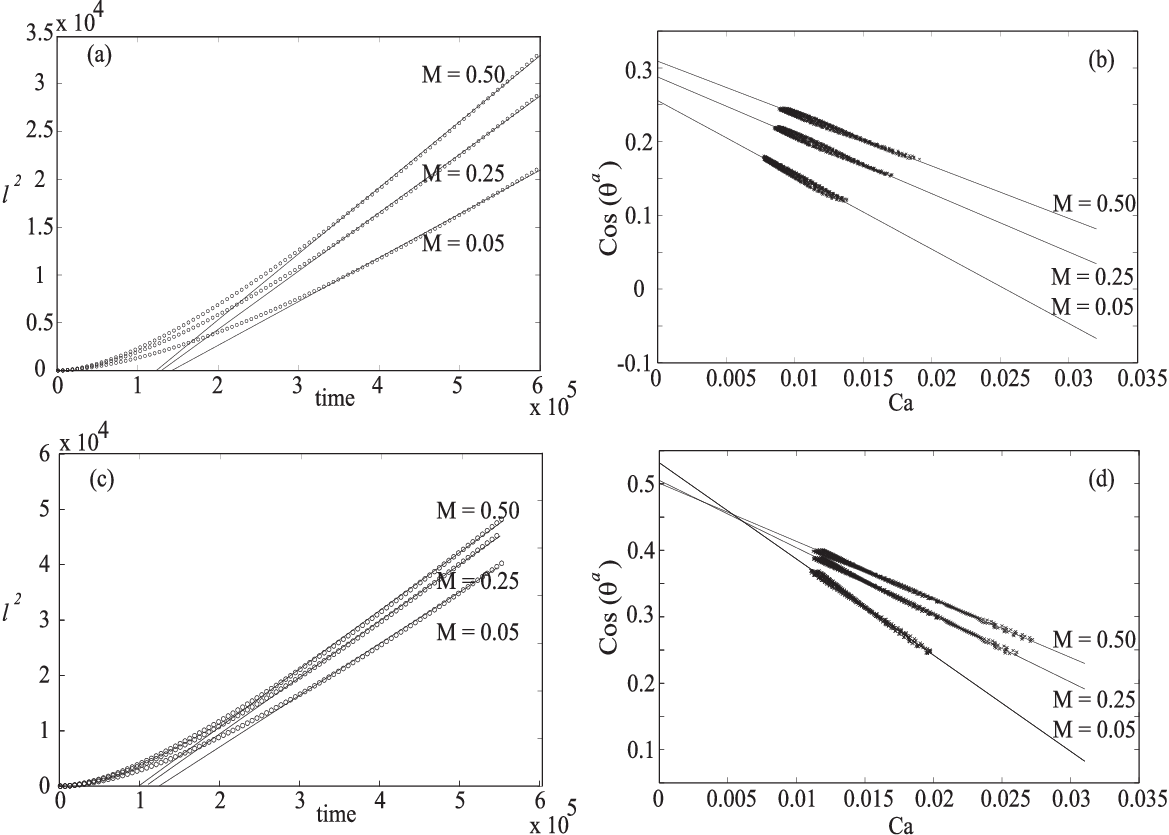}
\end{center}
\caption{Lattice Boltzmann simulation results for capillary filling in smooth channels. Left:
the length squared, in lattice units, of the column of the filling component (fluid $\alpha$) plotted against time for (a)
single and (c) multiple-relaxation-time lattice Boltzmann simulations. The circles are the simulation 
results and the solid lines are fits to Washburn's law. Right: the advancing 
contact angle of the liquid-liquid interface for (b) single and (d) multiple-relaxation-time simulations. The crosses are the simulation results and the solid
lines are linear fits of $\cos{\theta^a}$ to the capillary number \cite{Rothman}.}
\label{washburnfig}
\end{figure*}

Numerical results showing capillary filling of a two dimensional capillary are presented in 
Fig.~\ref{washburnfig} for both the single and the multiple-relaxation-time lattice Boltzmann algorithms. 
The simulations are for a channel of length $L=640$ and width $h=50$. Reservoirs 
($480 \times 200$) of the two components are attached at each end of the capillary. 
The two reservoirs are connected to ensure that they have the same pressure. 
The other parameters of the model are chosen to give $\theta^{eq} = 60^{\mathrm{o}}$, 
$\gamma = 0.0188$, and $\rho_{\mathrm{1}} = \rho_{\mathrm{2}} = 1.0$. Fluid $\alpha$, with viscosity $\nu_{\mathrm{1}} = 0.83$ is taken to displace fluid $\beta$ with viscosity $\nu_{\mathrm{2}} = 0.03$. Results are shown for mobilities $M = 0.05$, $M=0.1$ and $M=0.5$.

The solid lines in Fig.~\ref{washburnfig}(a) and (c) are fits to the Washburn law. 
At later times, once inertial effects have become negligible, both the single and multiple-relaxation-time lattice Boltzmann simulations satisfy the Washburn scaling. However,
the two methods provide different quantitative results, in particular, capillary filling is 
considerably slower for the single-relaxation-time method. The difference can be 
explained by considering how the advancing contact angle varies with the velocity of the interface, $v^{I}$. 

To lowest order in the capillary number, Ca$=v^{I} \nu/\gamma$, the advancing contact angle is related to the equilibrium angle and the capillary number by \cite{Rothman}
\begin{equation}
\cos{\theta^{a}} = \cos{\theta^{eq}} - \mathrm{Ca} \, \mathrm{log} (KL/l_s),
\end{equation}
where $K$ is a constant, $L$ is the length scale of the system and $l_s$ is the effective slip length 
at the three phase contact line. 
Figs.~\ref{washburnfig}(b) and (d) show the expected linear decrease of the measured contact angle with capillary number \cite{Rothman}. 
For the multiple-relaxation-time algorithm, the advancing contact angle tends to the correct value as $\mathrm{Ca}\rightarrow 0$. 
We obtain $\theta_a |_{Ca\rightarrow 0}=58^{\mathrm{o}}$, $60^{\mathrm{o}}$ and $60^{\mathrm{o}}$  
for $M=0.05$, $0.1$ and $0.5$ respectively. For the BGK method, however, $\theta^a |_{Ca\rightarrow 0}=75^{\mathrm{o}}$,
$73^{\mathrm{o}}$ and $72^{\mathrm{o}}$ for $M=0.05$, $0.1$ and $0.5$, and the advancing angle is consistently 
higher for all values of Ca than for the multi-relaxation-time solution. (This result agrees with that presented in Fig.~\ref{fig2}, 
which shows that  the measured equilibrium contact angle is too high in the single-relaxation-time approach.) Since the speed of capillary filling depends 
on $\cos{\theta^a}$ (\ref{washburnlaw}), capillary filling is considerably slower using this method. 

In diffuse interface models of binary fluids the contact line singularity is relieved by inter-diffusion of the two fluid components. This is governed by the mobility $M$; as is apparent from Fig.~\ref{washburnfig}, increasing $M$ increases the rate of diffusion and hence the velocity of the contact line. Therefore the parameter $M$ can be used to tune the effective slip length  $l_s$. 


\section{Summary and Conclusions}
\label{summary}

We have shown that, if the lattice Boltzmann relaxation parameter $\tau_{\rho} \neq 1$, 
strong spurious currents drive the contact line out of 
equilibrium even in a system at rest. This means 
that the algorithm gives incorrect values for the contact angle. In many 
applications it is possible to choose $\tau_{\rho}=1$, thus avoiding the problem. 
However this parameter controls the fluid viscosity and so cannot be held at unity 
for both phases of a binary fluid if the two components have different viscosities.

We demonstrate that the spurious currents arise primarily from two effects. The first is 
a long-range contribution to the equilibrium distribution function near the contact 
line that effectively originates in the incorrect phase. The second is the 
bounce-back boundary conditions which drive the interface out of equilibrium. 

Aiming to reduce the unwanted velocities we propose a revised lattice Boltzmann 
method, based on a multiple-relaxation-time algorithm. We show that the simulations 
then agree well with the analytical result for the equilibrium contact angle.  
Moreover the dynamic, advancing contact angle shows the expected physical behaviour, 
with a slip length that depends on the diffusivity of the fluid. 

Using this method it will be possible to perform accurate simulations of a binary fluid where the two components have different viscosities. Examples of problems where the algorithm will prove useful include viscous fingering, rolling of viscous drops and capillary filling. Moreover, it can provide a useful approximation to a liquid-gas system in the limit that evaporation-condensation is not important.

\appendix

\section{The choice equilibrium distribution}
\label{app1}

We list the best choice of equilibrium distributions and stencils for calculating  
spatial derivatives for the lattice Boltzmann algorithms we have considered in this paper, 
based on reducing the magnitude of spurious velocities near to interfaces (a  
detailed account is given in \cite{kalli}).  

The equilibrium distributions can be written in the form
\begin{eqnarray}
f^{eq}_i({\bf r}) &=& \tfrac{w_{i}}{c^{2}} \Big( p_0 - \kappa \phi \nabla^2 \phi +  
e_{i\alpha} \rho u_{\alpha} 
 \nonumber\\
&&
\hspace{-0.2cm}
 \quad + \tfrac{3}{2c^{2}} \left[ e_{i\alpha} e_{i\beta} - \tfrac{c^{2}}{3}\delta_{\alpha\beta}\right]  \rho u_\alpha u_\beta \Big)  
\nonumber\\
&&
\hspace{-0.2cm}
\quad + \tfrac{\kappa}{c^2} \Big( w_i^{xx}  \partial_x \phi \partial_x \phi \! +\! w_i^{yy}  
\partial_y \phi \partial_y \phi \!+\! w_i^{xy} \partial_x \phi \partial_y \phi \Big), \nonumber\\
g^{eq}_i({\bf r}) \!&=&\! \tfrac{w_{i}}{c^{2}} \Big( \Gamma \mu  \!+\!
e_{i\alpha} \phi u_{\alpha} 
\!+\! \tfrac{3}{2c^{2}} \left[ e_{i\alpha} e_{i\beta} -
\tfrac{c^{2}}{3}\delta_{\alpha\beta}\right] \phi u_\alpha u_\beta  \Big), \nonumber\\
\nonumber  \\
\label{equilibrium}
\end{eqnarray}
for $i=1,..,8$, where $w_{1\text{-}4} = \tfrac{1}{3}$, $w_{5\text{-}8} = \tfrac{1}{12}$, and  
summation over repeated indices is assumed.
Other parameters are $w_{1\text{-}2}^{xx} = w_{3\text{-}4}^{yy} = \tfrac{1}{3}$,  
$w_{3\text{-}4}^{xx} = w_{1\text{-}2}^{yy} = -\tfrac{1}{6}$, $w_{5\text{-}8}^{xx} =  
w_{5\text{-}8}^{yy} = -\tfrac{1}{24}$, $w_{1\text{-}4}^{xy} = 0$, $w_{5,6}^{xy} = \tfrac{1}{4}$, and $w_{7,8}^{xy} = -\tfrac{1}{4}$.

%
%

The $i=0$ stationary values are chosen to conserve the concentration of each species:
\begin{eqnarray}
f_0^{eq}({\bf r}) = \rho - \sum_{i=1}^{8} f_i^{eq}({\bf r}), \quad   
g_0^{eq}({\bf r}) = \phi - \sum_{i=1}^{8} g_i^{eq}({\bf r}).
\label{eqn:feq0}
\end{eqnarray}

During the lattice Boltzmann procedure, it is necessary to calculate numerically both derivatives ({\it e.g.} $\partial_x \rho$ in the equilibrium distribution (\ref{equilibrium})) and the Laplacian ({\it e.g.} to obtain the chemical potential (\ref{chempot})). These continuous quantities are calculated from stencils, discrete operators which use neighbouring lattice sites. 
The best choice of stencils to reduce spurious velocities is given by \cite{kalli}  
\begin{eqnarray}
\bar{\partial}_x \! = \!
\tfrac{1}{12 \Delta x} \!
\left[
\begin{array}{ccc}
-1 & 0 & 1 \\
-4 & 0 & 4 \\
-1 & 0 & 1 \\
\end{array}
\right] \!
\!,  
\bar{\nabla}^2  \! = \!
\tfrac{1}{6 {(\Delta x})^2} \!
\left[
\begin{array}{ccc}
1 & 4 & 1 \\
4 & -20 & 4 \\
1 & 4 & 1 \\
\end{array}
\right] \!\! .
\label{best}
\end{eqnarray}
The values in these matrices denote the weights given to a particular quantity on a lattice node (the central entry) and on the surrounding eight lattice points.


\begin{thebibliography}{99}
\bibitem{tau} To eliminate spurious long-range effects we always set $\tau_\phi = 1$. This, however, is not restrictive because the model contains another parameter, $\Gamma$, which can be used to alter the particle mobility.
\end{thebibliography}
\end{document}